\documentclass[aps,12pt,preprint]{revtex4}
\usepackage{epsfig}
     \font\tenbifull=cmmib10 scaled 1200 
     \font\tenbimed=cmmib9
      \font\tenbismall=cmmib7
\mathchardef\bbkappa="7114 \mathchardef\bbrho="711A
\mathchardef\bbsigma="711B \mathchardef\bbtau="711C
\mathchardef\bbvarrho="7125 \mathchardef\bbvarsigma="7126
\mathchardef\bbPhi="7008 \mathchardef\bbxi="7118

\newcommand {\CM} {{\cal M}}

\newcommand{\be}{\begin{eqnarray}&&}
\newcommand{\ee}{\end{eqnarray}}
  \usepackage{graphicx}
\begin{document}
\title{$\omega$ and $\eta$ ($\eta$') mesons from $NN$ and $ND$ collisions at intermediate
energies}

\author{L. P. Kaptari}
\affiliation{Bogoliubov Laboratory of Theoretical Physics, 141980, JINR, Dubna,
Russia}
\affiliation{Department of Physics, University of Perugia;
      Istituto Nazionale di Fisica Nucleare, Sezione di Perugia,
      Via A. Pascoli, I-06123, Italy}
\altaffiliation{through the program Rientro dei Cervelli of the
Italian Ministry of University and  Research}

\author{B. K\"ampfer}
\affiliation{Research Center Dresden-Rossendorf, 01314 Dresden, PF 510119, Germany\\
TU Dresden, Institut f\"ur Theoretische Physik 01062 Dresden,
Germany}

\begin{abstract}

The production of  pseudo scalar, $\eta$, $\eta'$, and vector,
$\omega$, $\rho$, $\phi$,  mesons  in NN collisions at
threshold-near energies is analyzed within a covariant effective
meson-nucleon theory. It is shown that  a good description of
cross sections and angular distributions, for vector meson
production, can be  accomplished by considering meson and nucleon
currents only, while for pseudo scalar production an inclusion of
nucleon resonances is needed.
 The  di-electron production  from subsequent   Dalitz decay of the produced mesons,
  $\eta'\to  \gamma \gamma^* \to\gamma e^+e^-$ and $\omega\to
  \pi\gamma^*\to \pi e^+e^-$
  is also considered and numerical results are presented for intermediate energies and
 kinematics  of possible experiments with HADES, CLAS and KEK-PS.
 We argue that  the transition form factor $\omega\to \gamma^*\pi$
 as well as $\eta'\to \gamma^*\gamma$
 can be defined in a fairly model independent way
 and the feasibility of an experimental access to transition
form factors is discussed.
\end{abstract}

\maketitle
 {\bf I. Introduction}

A  theoretical analysis of light pseudoscalar and vector meson production in
 $pp\to ppM$,  $pn\to pnM$, $pn\to dM$  and  $dp\to p_{sp} \, np \, e^+e^-$ processes
(here $M$ denotes a meson, pseudoscalar $\eta$ and $\eta'$, or vector,   $\omega$ or $\phi$; $p$ ($n$)
denotes the proton  (neutron), $p_{sp}$ is the spectator proton, $d$ stands for the deuteron,
and $e^+e^-$ for a di-electron pair)
 at threshold-near energies is interesting for
different aspects of contemporary  particle and nuclear physics.
It is known that the effective $NN$ forces at
 short distances are governed by exchanges of $\rho$ and $\omega$ so that a
 study of their contribution to the $NN$ elastic amplitude  and
 to the Meson Exchange Currents  in elastic scattering from
 light nuclei (e.g., the deuteron) can substantially
 augment the knowledge of the short-range part of the potential.
 Another important issue is the   di-electron emission in $NN$ collisions which
 supplies  additional information on production of vector mesons
 with similar quantum numbers but rather different quark contents, in particular
 $\omega$ and $\phi$ mesons, which is interesting in respect  to
 the Okubo-Zweig-Iizuka (OZI) rule and study of hidden strangeness in the nucleon.
According  to the OZI rule   the
production of $\phi$ mesons in nucleon-nucleon collisions should
be strongly suppressed relative to $\omega$ production. An
enhanced $\phi$ production  would imply some exotic (e.g., hidden
strangeness) components in the nucleon wave function.

The pseudo-scalar mesons $\eta$ and $\eta'$ represent a subject of
considerable interest since some time
(cf.\ \cite{diekmann}
for reports). Investigations of various aspects $\eta$ and $\eta'$
mesons are tightly related with several theoretical challenges and
can augment the experimental information on different
phenomenological model parameters. Also,  near the threshold the
invariant mass of the $NN\eta'$  system in such reactions is in
the region of heavy nucleon resonances, i.e. resonances with
isospin $1/2$, including the so-called "missing resonances", can be investigated via these processes.
 Another aspect of $\eta$ and $\eta'$ production is that
 they constitute important  sources of di-electrons in $NN$ reactions.
It is, in particular, the $\eta$ which is significant source of
$e^+ e^-$ pairs, competing at invariant masses of 150 - 400 MeV
with $\Delta$ Dalitz decays and bremsstrahlung \cite{our_bremsstrahlung}, as the analysis
\cite{eta_contributions} of HADES data \cite{HADES} shows. One of
the primary aims of the HADES experiments \cite{HADES} is to seek
for signal of chiral symmetry restoration in compressed nuclear
matter. For such an endeavor one needs a good control of the
background processes, including the  Dalitz decay, in
particular at higher beam energies, as becoming accessible at
SIS100 within the FAIR project \cite{FAIR}.
The   Dalitz decays of mesons depend on the  transition,
 "vector-to-pseudoscalar" or "pseudoscalar-to-vector", form factors (FF)
 which encode hadronic information accessible in first-principle QCD calculations or
QCD sum rules. The Dalitz decay process of a  meson (pseudo scalar
$"ps"$ or vector $"V"$) can be presented as $ ps (V) \to \gamma(ps)
+\gamma^* \to \gamma (ps)  + e^- + e^+. $ Obviously, the
probability of emitting a virtual photon is governed by the
dynamical electromagnetic structure of the "dressed" transition
vertex $ps \to \gamma \gamma^*$ ($V\to ps \, \gamma^*$) which is encoded in the transition
form factors. If the decaying particle were point like, then
calculations of mass distributions and decay widths would be
straightforwardly given by QED. Deviations of the measured
quantities from the QED predictions directly reflect the effects
of the form factors and thus the internal hadron structure.

For a reliable study of these effects one needs more experimental data
and more types of processes.  In particular, for further checks of
the reaction mechanism it is necessary to  study meson production
also at neutron targets which can be be extracted, with some
efforts and even mostly with some model dependent assumptions,
from reactions on nuclei, mainly on the deuteron.
 The spectator technique  represents one example how one can use
a deuteron target to isolate reactions on the neutron. It is based
on the idea to measure the spectator proton, $p_{\rm sp}$, at
fixed beam energy in the  meson  production reactions $ d \, p\to  p_{\rm sp} \,np M$,
thus exploiting the internal momentum spread
of the neutron inside the deuteron. In such a way one gets access
to quasi-free reactions $p n$.

 In the present paper we present a theoretical approach within that an analysis  of
 mentioned processes can be achieved on a common ground and the
 differential and total cross sections, as  functions of the relevant kinematical variables and
 initial energy,  can be parametrized with the same set of effective parameters.

{\bf II. The model}

\begin{figure}[h]  
 \includegraphics[width=0.85\textwidth]{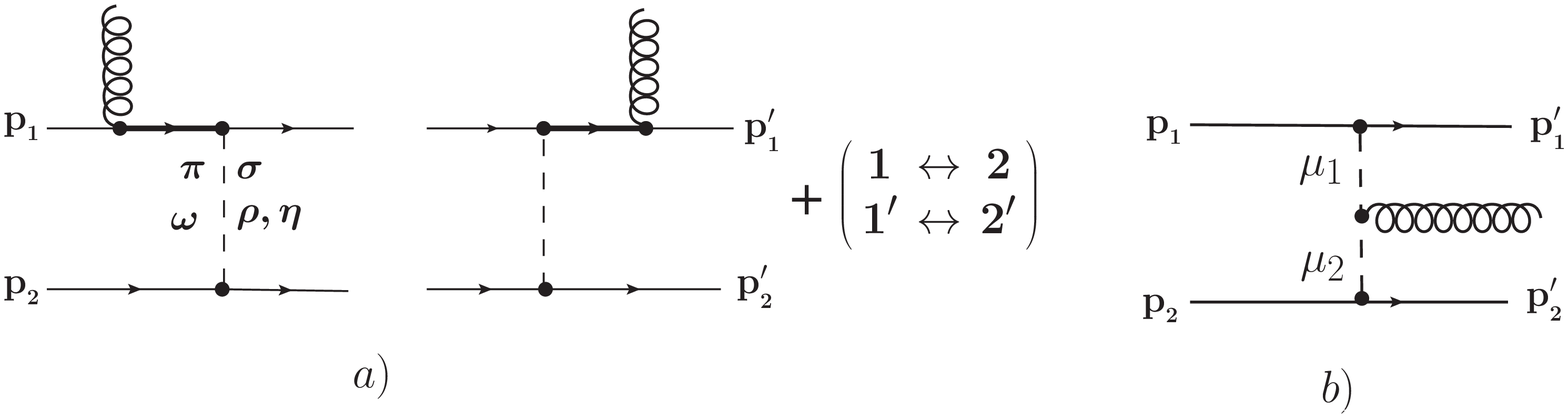} %
 \caption{Feynman diagrams for the nucleon current (a) and internal conversion (b)
 contributions to
meson production in $NN$ reactions. The thin  solid lines denote
incoming and outgoing nucleons, the dashed lines are for the
exchanged (OBE) mesons, while the intermediate thick lines can be either
a virtual nucleon or a nucleon resonance;   $\mu_1$ ($\mu_2$) are the virtual mesons
before (after) conversion, whereas the produced final meson is
depicted as waved lines.}
\label{fig1}
\end{figure}

The nucleons and  mesons,  involved in the  process are treated
 within a meson-nucleon  theory based on
 effective interaction Lagrangians  with
scalar, pseudoscalar,  and neutral   ($  \omega, \phi$) and
charged vector ($\rho$)  mesons (see e.g. \cite{our_omega_phi,our_bremsstrahlung,Nakayama}).
The electromagnetic interaction Lagrangians are included into the model as well.
The invariant   cross section for the  meson production in $NN$
collisions of the type $ N_1+N_2 \to N_1'+N_2'+ps(V)$ is
\begin{eqnarray}
&& d^5\sigma = \frac{1}{2\sqrt{s(s-4m^2)}} \frac14
\sum\limits_{s_1,s_2}\ \sum\limits_{s_1',s_2',\CM_V}
\,|T_{s_1s_2,s_1',s_2'}^{\CM_V } |^2 d^5\tau_f \ \frac{1}{n!
}, \label{crossnn}
\end{eqnarray}
where $m$ is the nucleon mass,    $s_i$ and $\CM_V$ are the projections of the nucleonic and
mesonic spins ($\CM_{ps}=0$) on the quantization axis, $d\tau_f$ is the
invariant phase space volume, $s$ is the invariant mass squared of the initial particles and the factor
$\displaystyle\frac{1}{n!}$ accounts for $n$ identical particles
in the final state.
Calculations of the amplitude $T_{s_1s_2,s_1',s_2'}^{\CM_V }$ with the chosen Lagrangians
 result into a series of Feynman diagrams of  two types:
 (i) the ones which describe
the meson production from the processes of one-boson exchange
(OBE) between two nucleons accompanied by the emission   of a
meson by a nucleon
 (in what
follows we call these diagrams nucleon current contribution, see
Fig. \ref{fig1}a), and (ii) production of  mesons resulting from a
conversion of  virtual
 exchanges into a  meson, which are called
internal  conversion type diagrams, Fig. \ref{fig1}b).

In our calculations we use the following Lagrangians:
(i) Nucleon currents:
\begin{eqnarray}&&
{\cal L}_{\sigma NN }= g_{\sigma NN} \bar \Psi_N(x)\Psi_N(x) \it\Phi_\sigma(x) ,\label{unu}\\
&&
{\cal L}_{a_0 NN }= g_{a_0 NN} \bar\Psi_N(x) (\tau \Phi_{a_0})(x) \Psi_N(x)\it ,\\
&& {\cal L}_{ps NN}= -\frac{f_{ ps NN}}{m_{ps}}\bar
\Psi_N(x)\gamma_5\gamma^\mu \partial_\mu ({
\Phi_{ps}(x)})\Psi_N(x) , \\ &&
\!\!\!\!\!\!\!\!\!\! {\cal L}_{V NN}= -g_{V NN }\bar \Psi_N(x)
\left( \gamma_\mu {\Phi_ V}^\mu(x)-\frac{\kappa_V}{2m}
  \sigma_{\mu\nu} \partial^\nu{\Phi_V}^\mu(x)\right)\Psi_N(x)
\end{eqnarray}
(ii) Spin $\frac12 $ resonances ($S_{11}$ and $P_{11}$):
\begin{eqnarray}
{\cal L}_{NN^*ps}^{(\pm)}(x)&=&\mp \frac{g_{NN^* ps} }{m_{N^*}
\pm m_N}
 \bar\Psi_R(x)\left\{
\begin{array}{c}
\gamma_5\\ 1\end{array} \right\} \gamma_\mu\partial^\mu \Phi_{ps}(x) \Psi_N(x) + h.c. \\
{\cal L}_{NN^*V}^{(\pm)}(x)&=&  \frac{g_{NN^* V} }{2(m_{N^*}  +
m_N) }
 \bar\Psi_R(x)\left\{
\begin{array}{c}
1\\ \gamma_5 \end{array} \right\} \sigma_{\mu\nu} V^{\mu\nu}(x) \Psi_N(x) + h.c. \label{rr}
\end{eqnarray}
with the abbreviations $ps \equiv \pi$ or $\eta$ or $\eta$',
$\Phi_{ps} \equiv( \tau \Phi_\pi(x)) $ or $\Phi_{\eta'}(x)$, $V
\equiv V_\omega(x)$ or $V(\tau\rho(x))$, and
$V^{\alpha\beta}=\partial^\beta V^\alpha - \partial^\alpha
V^\beta$.
Similar expressions hold for the
spin $\frac32 $ resonances ($D_{13}$ and $P_{13}$) which contribute mainly to the $\eta'$
meson production.
Furthermore needed for internal conversion interactions, such as
${\cal L}_{\rho\pi\, V}$
${\cal L}_{psVV}$, ${\cal L}_{\gamma ll }$,
and ${\cal L}_{ps \gamma\gamma}$ are listed in
\cite{our_eta_prime}.
All the  the nucleon -- nucleon (resonance) -- meson vertices are
dressed with cut-off form factors of the form reported in
\cite{our_eta,our_eta_prime}.

{\bf III. Results}

These seemingly many ingredients (coupling strengths, form factors
and their cut-offs, see \cite{our_eta,our_eta_prime}) may cause
the impression that the one-boson exchange approach to hadronic
observables contains too many free parameters
and  does not have too much predictive power.
However, a bulk of these apparently free parameters
are constrained by independent experiments and can be fixed from independent data, e.g. from
fitting the  elastic $NN$ phase shifts or from known decay widths of mesons
into different partial channels etc. Details of fixing parameters can be found, e.g. in
Refs. \cite{our_bremsstrahlung,Nakayama,our_eta_prime,our_eta}.

\begin{figure}[ht]  
\includegraphics[width=.45\textwidth]{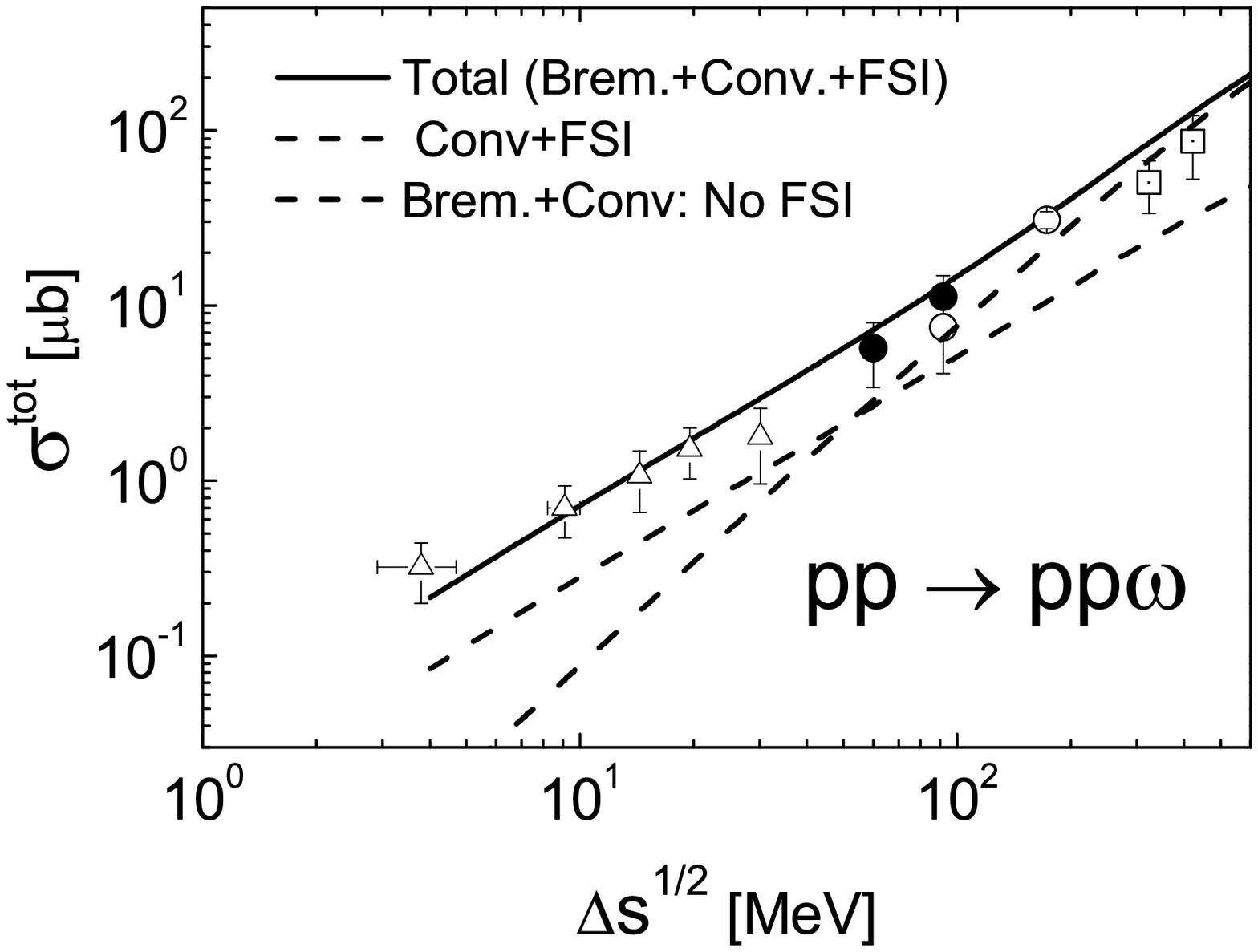} %
\hfill\includegraphics[width=.45\textwidth]{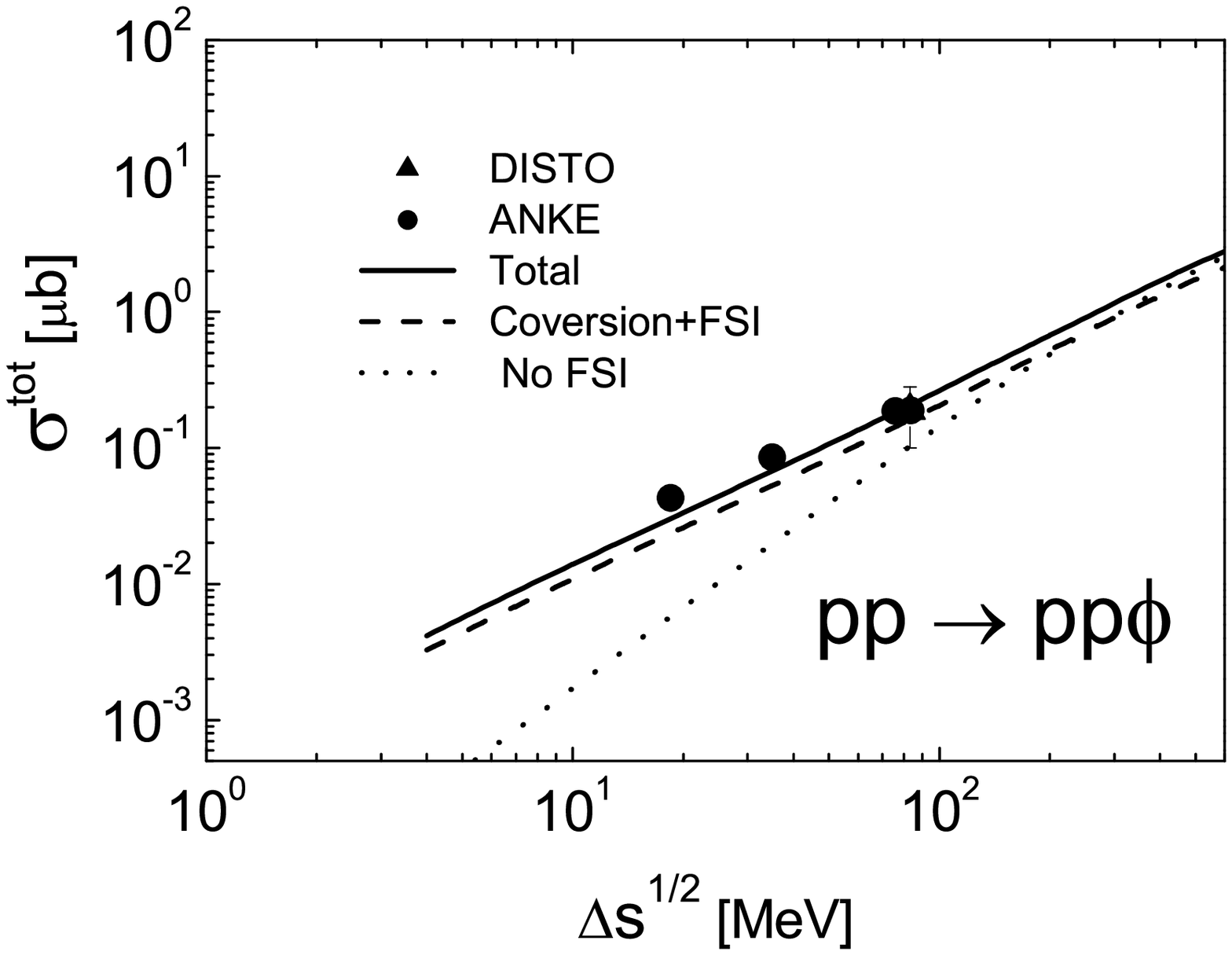} %
\includegraphics[width=.35\textwidth]{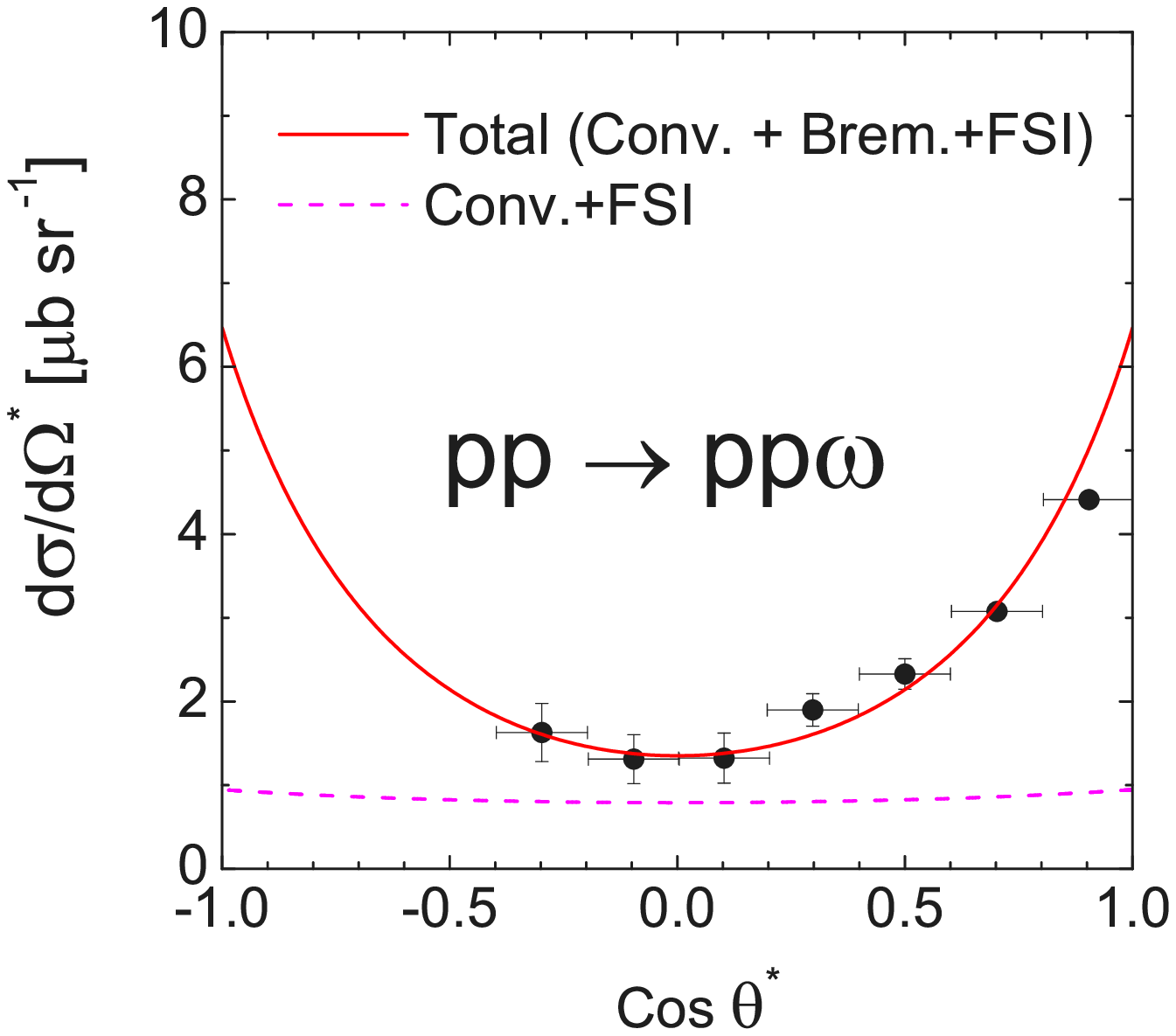} %
\hspace*{3cm}\includegraphics[width=.35\textwidth]{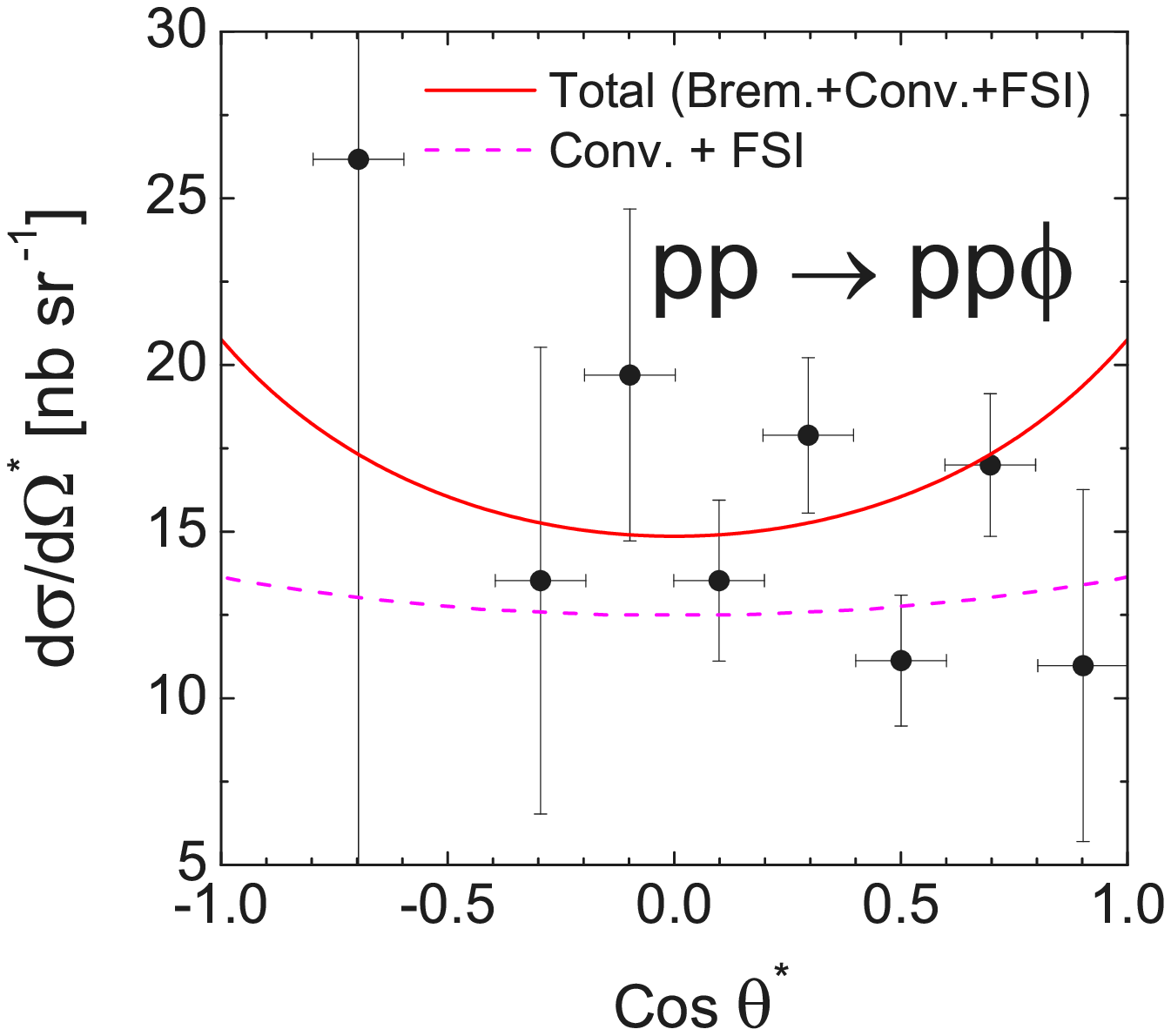}
\caption{Cross sections for $\omega$ (left) and $\phi$ (right)
production from \cite{our_omega,our_omega_phi}.  Experimental
data for $\omega$ are from \cite{TOF,ankenew} (open circles), \cite{hibou}
(triangles) and \cite{disto1} (squares), while for $\phi$  from
\cite{disto,disto1,newphi}. } \label{omegaphi}
\end{figure}

\begin{figure}[h]  
\vskip -3mm
\includegraphics[width=.5\textwidth]{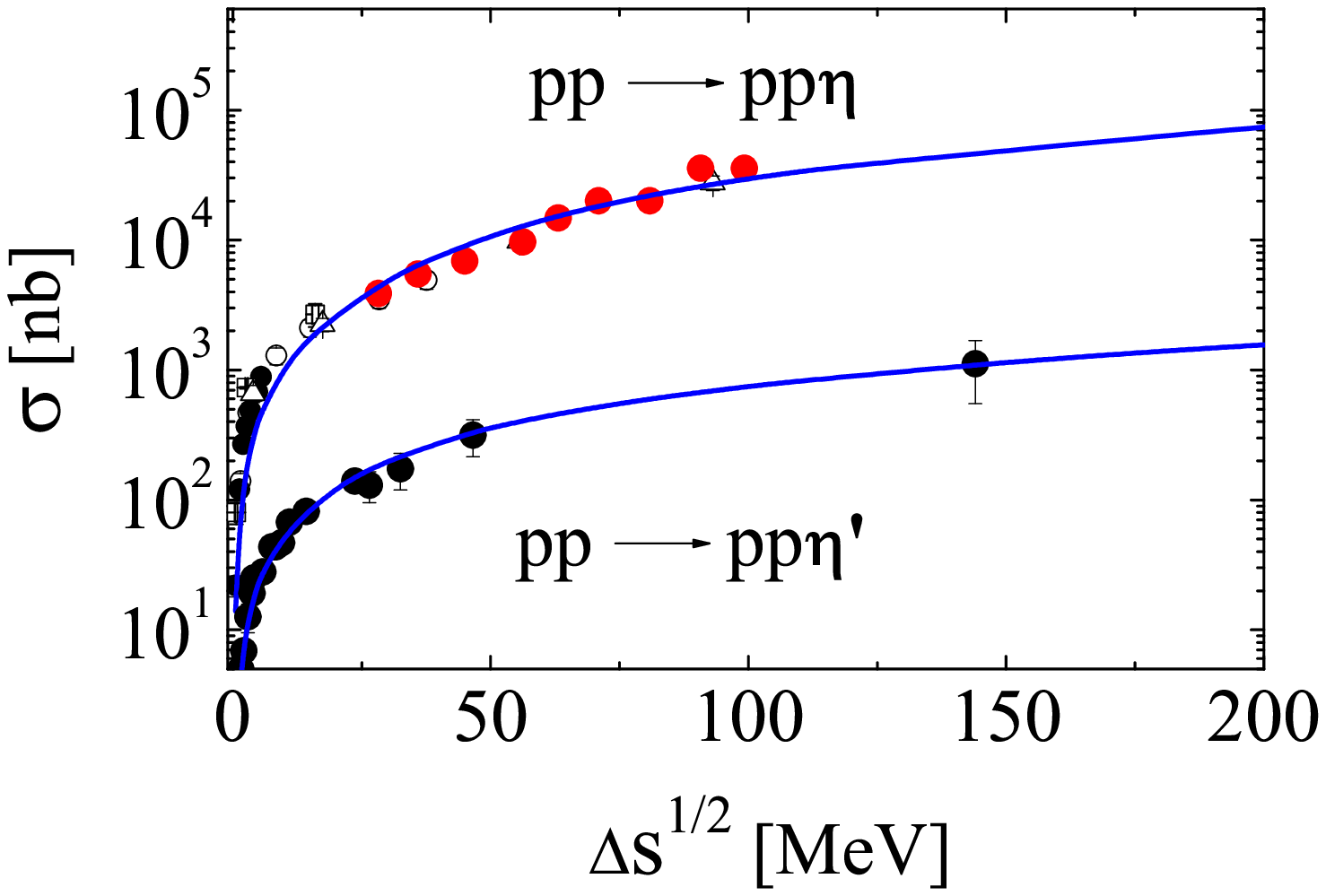} %
\includegraphics[width=.38\textwidth]{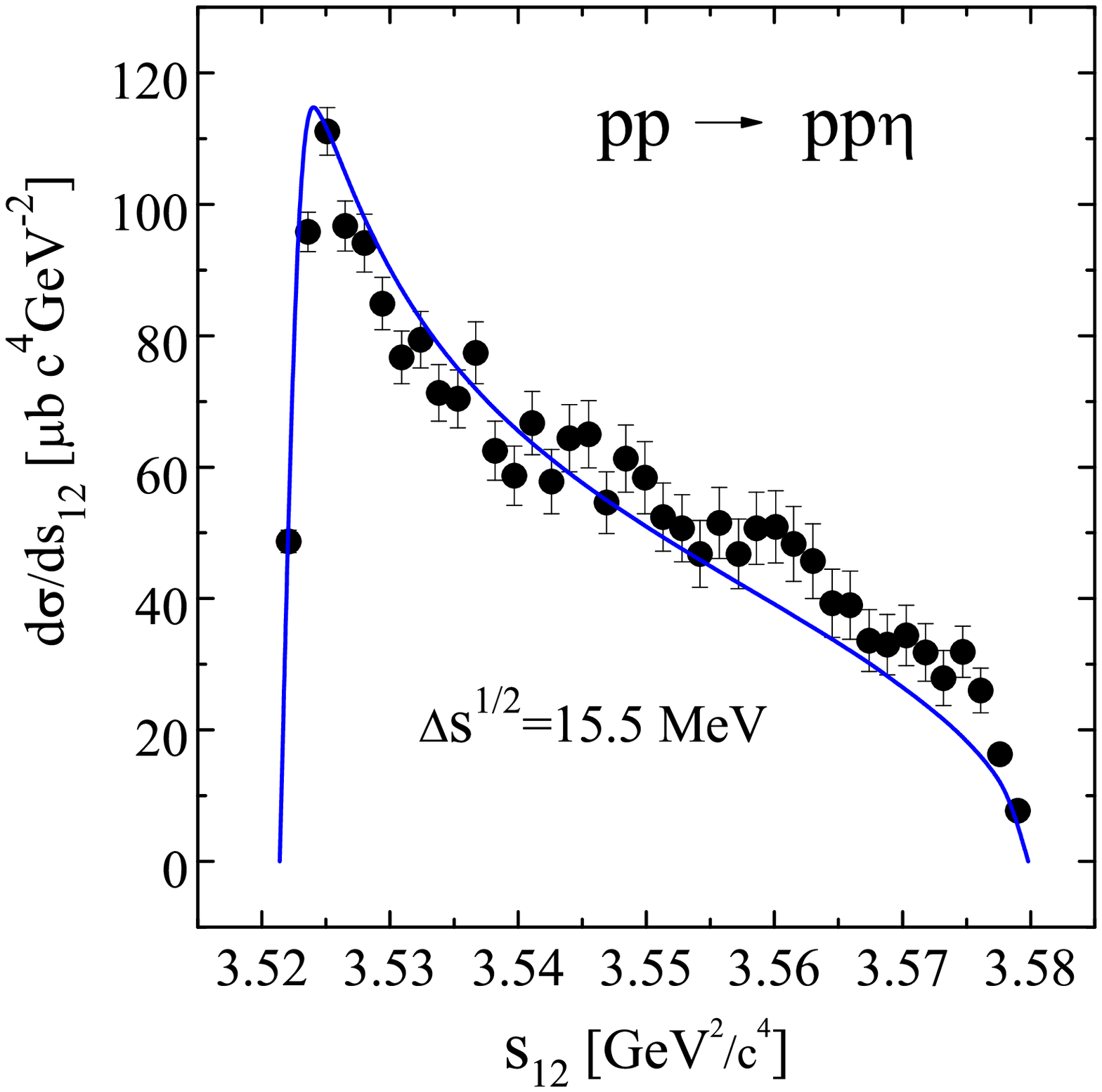} %
\vskip -1mm
\caption{Total cross sections for $\eta$  and
$\eta$'  production as a function of the energy excess in
$p+p$ (left) and invariant mass distribution of $\eta$ production as a function of the
invariant mass $s_{12}$ of the outgoing nucleons (right).  For data quotation
consult \cite{our_eta,our_eta_prime,eta,eta-prime}.}
\label{eta}
\end{figure}

\begin{figure}[h]
\includegraphics[width=.43\textwidth]{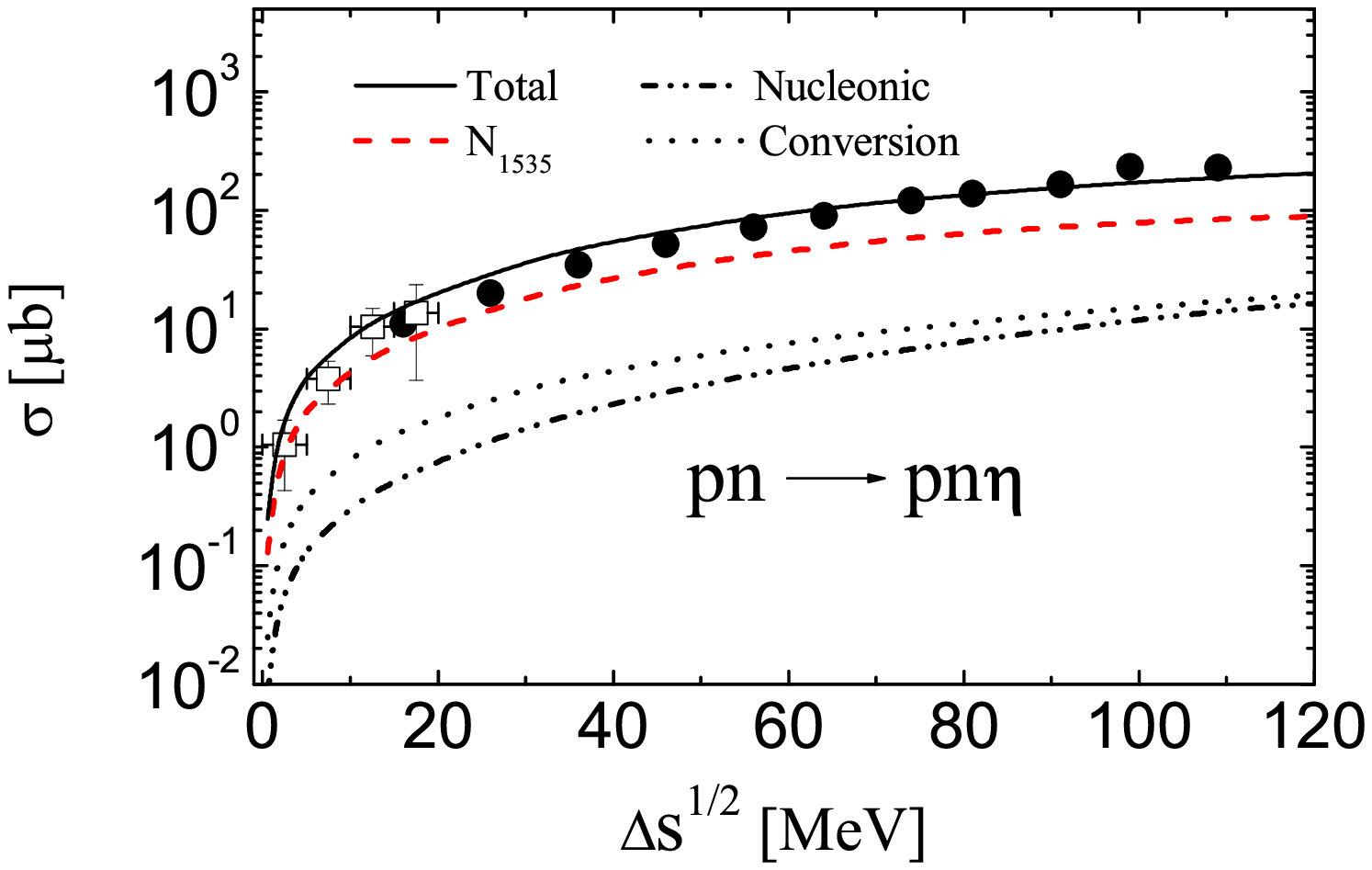} %
\hspace*{-4mm}
\includegraphics[width=.43\textwidth]{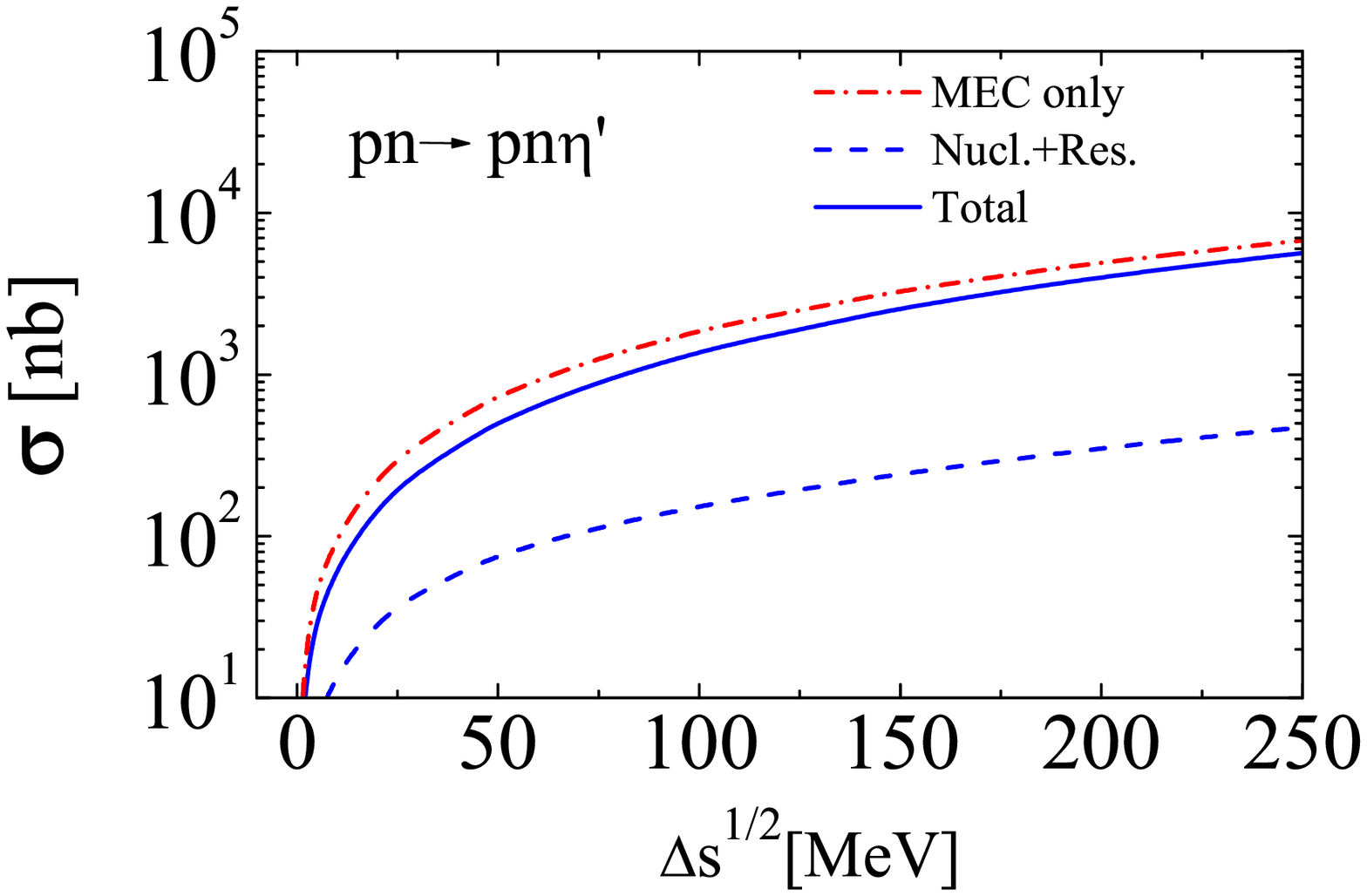} %
\vskip -7mm
\caption{.
Total cross section for  $\eta$  and $\eta$' rpduction in $pn$ reactions.
Experimental data are quoted in \cite{our_eta,our_eta_prime,eta,eta-prime}; the most  recent
results for $\eta$ production in $pn$ reactions~\cite{neweta08} are
depicted as squares (left). The role of different
contributions, nucleonic current, internal conversion (MEC) and resonances, is also displayed.} \label{pn}
\end{figure}

In
Fig.~\ref{omegaphi} the  results of calcualtions of the total cross sections (upper panel)
and angular distributions (lower panel)
for vector meson production ($\omega$ and $\phi$) \cite{our_omega,our_omega_phi}
are exhibited together with available experimental data.
It should be stressed that an overall good description of data has been achieved
by taking into account contributions from nucleonic current and internal conversion only, without
implementing any excitations of nucleon resonances. Also, as input into the calculations we used
the coupled constants for free $NNV$ vertices which do not contradict the OZI rule. The obtained
relatively high cross sections for $\phi$ production demonstrate, that the observed enhancement
is solely governed by dynamic effects (OBE interaction, interference of many different diagrams,
isospin effects etc.) and does not favour any OZI rule  violation and the presence of hidden strangeness
in nucleons. In Figs.~\ref{eta}~and~\ref{pn} results of calculations of $\eta$ and $\eta'$ mesons in
$pp$ and $pn$ reactions are displayed.
Contrarily to the case of vector mesons, excitations of intermediate resonances here are rather important.
This mainly concerns $\eta$ meson production, which occurs  primarily due to excitations of the $N_{1535}$ nucleon resonance.

Further
applications of the present approach to $\omega$ and $\phi$
production involving a final deuteron, including polarization
observables, have been presented in \cite{our_deuteron_phi}, while
\cite{our_bremsstrahlung} extends the formalism to virtual
bremsstrahlung in $NN \to NN \gamma^* \to NN e^+ e^-$ reactions.

 {\bf IV. Dalitz decay and transition form factors}

The subsequent Dalitz decay of the produced (virtual) meson into a di-electron pair and another
particle can be also described within the presented approach.
In the tree-level approximation the process $N_1+N_2\to N_1'+N_2'+ ps (V) \to N_1'+N_2'+ \gamma(ps)
+\gamma^* \to N_1'+N_2'+ \gamma (ps)  + e^- + e^+ $ is described by the same set of
 Feynman diagrams as in Fig.~\ref{fig1},
except that  now the produced meson (waved lines) is virtual and decays into the
considered channel.
The corresponding cross section then reads as

\begin{eqnarray}&&
\frac{d^2\sigma}{d s_{\gamma^*} ds_{M}}=
\sigma^{tot}\left (NN\to NN \, M \right)\,
\frac{\sqrt{s_{M}}/\pi}{\left ( s_{M} -M^2\right)^2}\,
\frac{d\Gamma_{[ps(V)\to \gamma (ps)e^+e^-]}}{ds_{\gamma^*}},
\label{twostep}
\end{eqnarray}

\begin{eqnarray}
\frac{d\Gamma_{[ps(V)\to \gamma (ps)e^+e^-]}}{ds_{\gamma^*}} =
\xi\displaystyle\frac{\alpha_{em}}{3\pi s_\gamma}
\frac{\lambda^{3/2}(s_M,s_\gamma,\mu_f^2)}{\lambda^{3/2}(s_M,0,\mu_f^2)}
\Gamma_{ps(V)\to  \gamma  \gamma(ps)}  \left | F_{ ps(V)  \gamma^* \,\gamma(ps) }(s_{\gamma^*})\right |^2,
\label{dgamma}
\end{eqnarray}
where
$s_M$ is the square of the invariant  mass  of the produced off-mass shell
meson with  $M=\mu-i\Gamma^{tot}/2$ as its pole mass and total decay width $\Gamma^{tot}$;
 $\mu_f^2=0,\, \xi=2$ in case of pseudoscalar Dalitz decay
 and $\mu_f^2=\mu_\pi^2, \xi=1$ for Dalitz decay
of a vector  meson ($\omega$).
The electromagnetic form factors  encode
non-perturbative transition matrix elements
$F_{ps(V) \gamma^* \gamma (ps)}$ in (\ref{dgamma}), basically accessible within QCD.
Here, however, we contrast a few parameterizations: (i) so-called
QED form factor meaning a structure-less particle with $\left |
F_{ {\eta'}  \gamma \gamma^*\,}(s_{\gamma^*})\right |^2 = 1$, (ii)
a parametrization suggested by the vector meson dominance (VDM)
model
\begin{equation}
F^{VMD}_{ {ps(V)} \gamma^*\gamma(ps)  \,}(s_{\gamma^*})=
\sum_{V=\rho,\omega,\phi} C_V \frac{m_V^2}{\hat
m_V^2-s_{\gamma^*}}, \label{fvmd}
\end{equation}
with $F(s_{\gamma^*}=0)=1$, $\sum C_V =
1$ and $\hat m_V = m_V - i \Gamma_V/2$. The values of $C_V$ are
quoted in~\cite{our_eta_prime}. For the case of light mesons ($\eta$ and $\omega$), the
kinematically accessible region is below the vector mesons pole masses and, as a
consequence, the $\rho$ contribution is sufficient. (iii) For
$\eta$', a monopole fit $ F_{\eta' \gamma \gamma^* }(Q^2) =
(1-Q^2/\Lambda_{\eta'})^{-1}$ \cite{our_eta_prime} may be used,
which does not differ too much from the VDM parametrization which, in this case,
includes $\omega$, $\rho$ and $\phi$ mesons.

\begin{figure}[h]  
\vskip -3mm
\includegraphics[width=.45\textwidth]{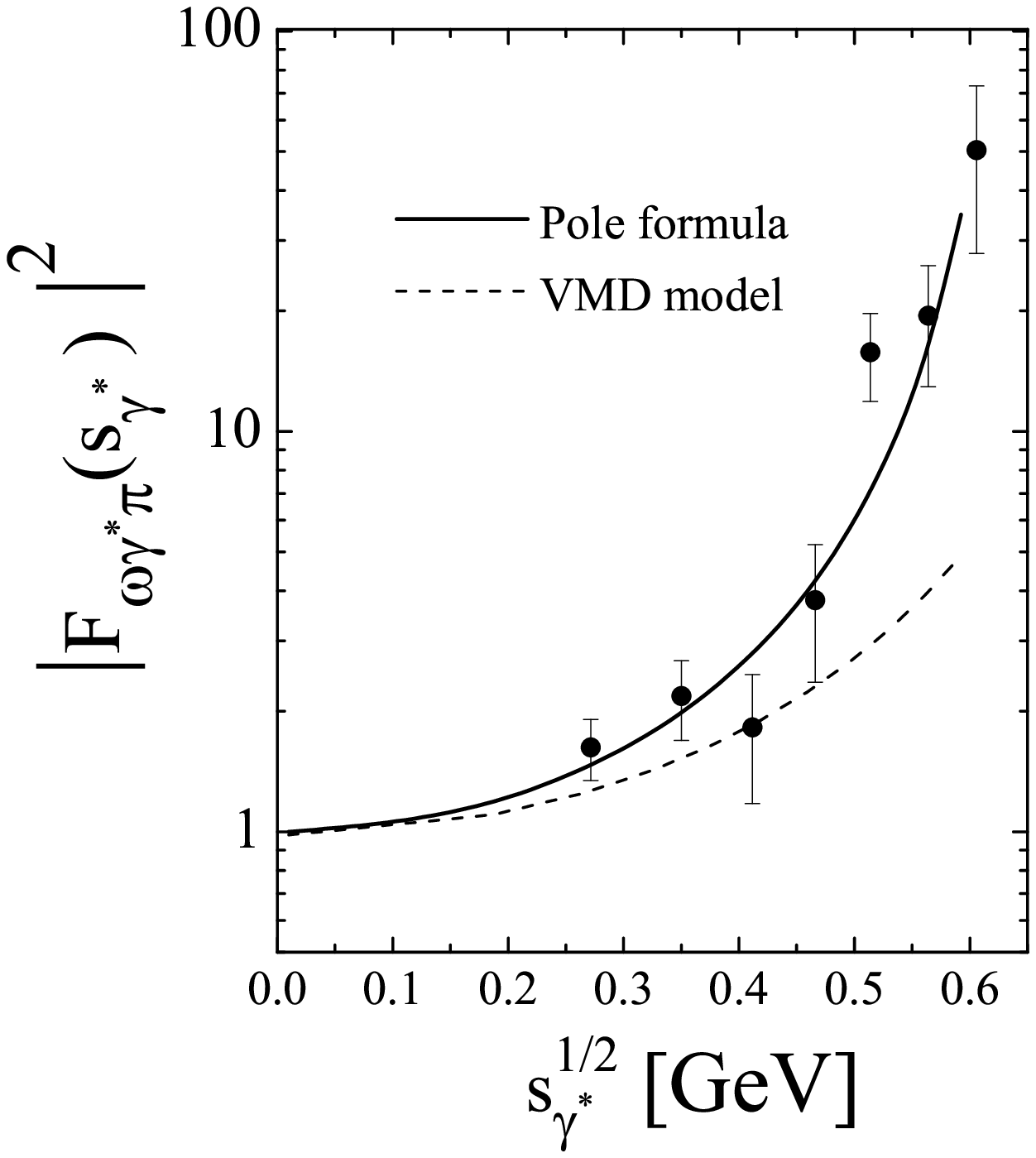} %
\includegraphics[width=.42\textwidth]{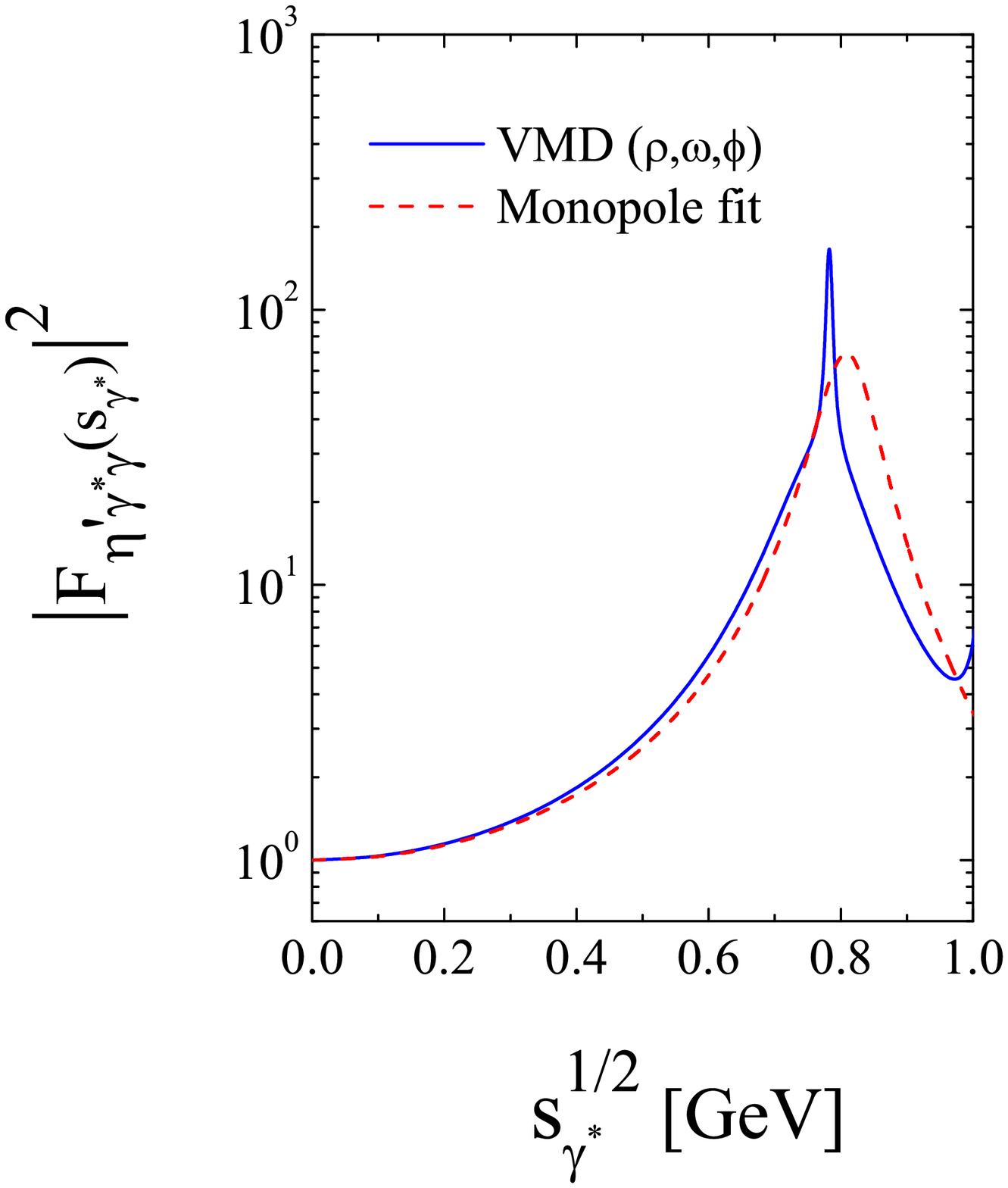} %
\vskip -3mm \caption{Transition form factors for Dalitz decay of $\omega\to \pi \, e^+e^-$ (left)
and $\eta'\to \gamma \, e^+e^-$ (right) }.\label{FF}
\end{figure}
In Fig.~\ref{FF} we present results of calculations of the  form factors defined by
eq. (\ref{dgamma}) for
transitions of  a vector mesons ($\omega$) into
a pion and a di-electron pair (left) and transitions of $\eta'$ into a real photon and  a di-electron pair (right).
The solid lines correspond to the VDM calculations, eq.~(\ref{fvmd}), while the
phenomenological fit is presented by the  dashed lines. It is seen, that since
for  $\omega$  decay the kinematically allowed values of
$s_{\gamma^*}$ (the argument of the transition FF) are below
 the vector meson pole masses, the corresponding FF exhibits a smooth behavior. A completely different situation
occurs in case of $\eta'$ meson, for which   $s_{\gamma^*}$  can be far beyond
the the vector meson pole masses and, correspondingly, the transition FF displays a sharp maximum near that poles.
Such a behavior can serve as a test of validity of VDM at large values of invariant masses.

{\bf VI. Reactions with the deuteron}

Eventually, herebelow we present the cross section of di-electron production in $d+p\to p_{sp}+n+p+e^+e^-$
reactions within the spectator mechanism, i.e. when the fast  proton is detected in forward direction with
the velocity not too different from the one of the incoming deuteron. The cross section is evaluated
 within the same effective meson nucleon theory, the deuteron now being treated within the Bethe-Salpeter
 formalism with the same effective parameters as in reactions with nucleons (details can be found in \cite{our_bremsstrahlung}).
\be
2E_{\rm sp}\frac{d\sigma}{d^3{\bf p}_{\rm sp}d s_{\gamma^*}}=2M_D\sqrt{\displaystyle
\frac{\lambda(s_{NN},m^2,m^2))}{\lambda(s_0,m^2,M_D^2)}}
n_D\left (|{\bf p}_{\rm sp}|\right)
\displaystyle\frac{d\sigma^{np}}{d s_{\gamma^*}},
\label{crossD}
\ee
where $n_D\left (|{\bf p}_{\rm sp}|\right)|$ is the deuteron momentum distribution in the deuteron center of mass system,
$s_0$ is the initial invariant energy of the colliding particles, $s_{NN}$ is the effective invariant energy of the
target proton and the neutron within the deuteron. It is seen that the desired cross section
$\frac{d\sigma^{np}}{d s_{\gamma^*}}$  at the neutron target
can be obtained from the experimentally measured cross section $2E_{\rm sp}\frac{d\sigma}{d^3{\bf p}_{\rm sp}d s_{\gamma^*}}$
by normalizing the latter with known kinematical factors $\lambda(s_{NN},m^2,m^2)$ and known deuteron momentum
distribution $n_D\left (|{\bf p}_{\rm sp}|\right)|$. Note, that eq. (\ref{crossD}) has been obtained strictly within
the spectator mechanism and can not be valid at large angles and/or low velocities of the spectator proton.

{\bf VII. Summary}

In summary we report on calculations of the reaction $NN \to NN \, M$ with $M$ as a pseudoscalar
$\eta, \eta'$ or vector $\omega,\rho, \phi$ meson and subsequent Dalitz decay of the produced
meson  within a one-boson exchange model. We point out
that isolating $\omega$, $\eta$ and $\eta$' contributions, e.g., in $p + p$
collisions, allows for an experimental determination of the corresponding
transition form factors. In particular,
for $\eta$' the vector meson dominance hypothesis would be testable.
On the other hand, the $\omega$ and $\eta$ Dalitz decay channel are
strong sources of $e^+ e-$ pairs in medium-energy heavy-ion
collisions which need to be understood before firm conclusions on
possible in-medium modifications of hadrons can be made. We
emphasize that, once the model parameters are adjusted in the
$p+p$ channel, the $n+p$ channel is accessible without further parameters.
The spectator technique with deuterons can allow for an experimental tagging of reactions at
the neutron target, provided the spectator proton is detected in the forward direction with
the same velocity as the incoming deuteron.

\end{document}